\documentclass{aa}
\usepackage{graphicx}

\newcommand{\lsun}{log$L/L_{\odot}\,$}
\newcommand{\msun}{$M_{\odot}\,$}

\begin{document}

   \title{Synthetic properties of bright metal-poor variables. I. ``Anomalous'' Cepheids.}

\author{G. Fiorentino \inst{1,2}, M. Limongi \inst{1}, F. Caputo \inst{1}, M. Marconi \inst{3}}

\institute{\bf{INAF-}Osservatorio Astronomico di Roma, Via
Frascati 33, 00040, Monte Porzio Catone,  Italy;
giuliana@mporzio.astro.it; caputo@mporzio.astro.it;
limongi@mporzio.astro.it \\ \and Universit\`a degli Studi di Roma
``Tor Vergata'', Via della Ricerca Scientifica 1,
00133, Roma \\
\and \bf{INAF-}Osservatorio Astronomico di Capodimonte, Via Moiariello 16,
80131 Napoli, Italy; marcella@na.astro.it }

\date{}

   \abstract{
We present new grids of evolutionary models for the so-colled
``Anomalous'' Cepheids (ACs), adopting $Z$=0.0001 and various
assumptions on the progenitor mass and mass-loss efficiency. These
computations are combined with the results of our previous set of
pulsation models and used to build synthetic populations of the
predicted pulsators as well as to provide a Mass-Luminosity relation
in the absence of mass-loss. We investigate the effect of mass-loss on
the predicted boundaries of the instability strip and we find that the
only significant dependence occurs in the Period-Magnitude plane,
where the synthetic distribution of the pulsators is, on average,
brighter by about 0.1 mag than the one in absence of mass-loss. Tight
Period-Magnitude relations are derived in the $K$ band for both
fundamental and first overtone pulsators, providing a useful tool for
distance evaluations with an intrinsic uncertainty of about 0.15 mag,
which decreases to $\sim$ 0.04 mag if the mass term is taken into
account. The constraints provided by the evolutionary models are used
to derive evolutionary (i.e, mass-independent) Period-Magnitude-Color
relations which provide distance determinations with a formal
uncertainty of the order of $\sim$ 0.1 mag, once the intrinsic colors
are well known.  We also use model computations from the literature to
investigate the effect of metal content both on the instability strip
and on the evolutionary Period-Magnitude-Color relations. Finally, we
compare our theoretical predictions with observed variables and we
confirm that a secure identification of actual ACs requires the
simultaneous information on period, magnitude and color, that also
provide constraints on the pulsation mode. }

\authorrunning{Fiorentino \it{et al.}}
\titlerunning{Synthetic properties of bright metal-poor variables. I. }

\maketitle
\pagebreak

\section{Introduction}

While the majority of metal-poor radial pulsating variables is
represented by the RR Lyrae stars, other classes of variables are
observed in metal-poor stellar systems.  According to the current
literature, one finds the ``Anomalous'' Cepheids (AC) and the
``Population II'' Cepheids (P2C), the former with periods ($P$) from
$\sim$ 0.5 to $\sim$ 2 days, the latter from $\sim$ 1 to 25 days. Both
types of variables are brighter than RR Lyrae stars and belong to the
central He-burning evolutionary phase, as are RR Lyrae stars. However.
ACs are more massive whereas P2Cs are less massive than RR Lyrae stars
with similar metal content. The purpose of the present paper is to
perform a theoretical analysis of the ACs, which are observed in the
majority of the Local Group dwarf galaxies which have been surveyed
for variable stars, but that are almost absent in other metal-poor
stellar systems such as Galactic Globular Clusters. The origin of this
``anomaly'' is the evidence that they do not follow the
Period-Luminosity relation of the P2Cs observed in these old metal
deficient clusters, being significantly brighter at fixed period (see
Norris \& Zinn 1975; Zinn \& Searle 1976; Smith \& Stryker 1986; Nemec
et al.  1994).

Several authors (Dolphin et al. 2002, 2003; Clementini et al. 2003;
Cordier, Goupil \& Lebreton 2003) have suggested that ACs are the
natural extension of the Population I Classical Cepheids to lower
metal contents and smaller masses. This suggestion is well supported
by theoretical investigations (Marconi, Fiorentino \& Caputo 2004
[MFC]; Caputo et al. 2004 [C04]) where, based on the constraints
provided by pulsation and evolutionary models, it is shown that
Anomalous and Classical Cepheids define a common region in the
$M_V$-log$P$ plane, with the former ones at lower luminosities and
shorter periods, as actually observed in the dwarf irregular galaxy
Leo A (Dolphin et al. 2002). Moreover, it is also shown that this
region is well separated from that populated by RR Lyrae stars and
P2Cs, in full agreement with observations.

Concerning the evolutionary phase, there is a general consensus that
ACs are central He-burning stars with a mass around 1.5\msun. Previous
studies (see Castellani \& Degl'Innocenti 1995 [CD95]; Bono et
al. 1997 [B97]; Caputo 1998; C04 and references therein) have shown
that for metal abundances $Z \le$ 0.0004 the central He-burning models
more massive than $\sim$ 1.2\msun evolve into the pulsation region at
a luminosity not dramatically higher than the RRL level, predicting
pulsators with AC-like periods and luminosities.

In this study, we will extend those investigations in order to
define a sound theoretical scenario for the analysis of the AC
properties. In particular, our intention is to verify their use as
distance indicators and to study the effects of mass-loss on the
various relations connecting evolutionary and pulsational quantities.

The paper is organized as follows: in Section 2, we present new
evolutionary tracks computed without mass-loss, as well as the central
He-burning models obtained from a given progenitor star assuming
different amounts of mass-loss. Using also the constraints provided by
pulsation models, synthetic populations of the predicted pulsators are
obtained. Section 3 presents the results for the Color-Magnitude
diagram as well as the Period-Magnitude, Period-Color, and
Period-Magnitude-Color relations. Section 4 deals with the comparison
between theoretical predictions and observations and the Conclusions
close the paper.

\begin{figure}
\includegraphics[width=8cm]{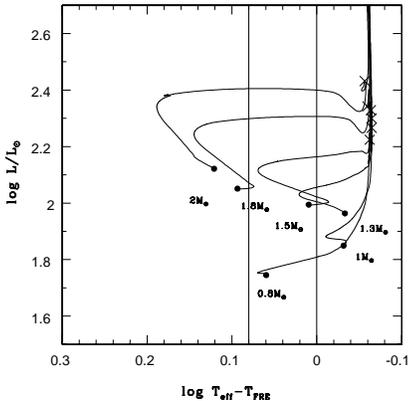}
\caption{\small{Canonical (no mass-loss) evolutionary tracks of
central He-burning models with $Z$=0.0001, $Y$=0.24 and the labeled
masses. In order to point out the connection of evolution with
pulsation, we subtract the value at the red edge (FRE) of the
pulsation region from the effective temperature of each model. Dots
indicate the initial central He-burning phase, crosses indicate
the central He-exhaustion phase corresponding to t$_{He}$ (see Table
1), as the vertical lines represent the predicted boundaries of
the instability strip (see text for details).}}
\end{figure}

\section{Evolutionary tracks}
\subsection{{\bf Canonical models}}

Our first step concerns the computation of ``canonical'' (i.e.,
without mass-loss) stellar models with mass $M$=0.8, 1.0, 1.3, 1.5,
1.8, and 2.0\msun at the chemical composition $Z$=0.0001 and
$Y$=0.24. The evolutionary tracks (available soon at the web site
http://www.mporzio.astro.it/~limongi/) have been computed from the
Pre-Main Sequence phase up to the end of the Early Asymptotic Giant
Branch (EAGB) by means of the latest versions of the FRANEC code,
whose main properties and input physics are extensively presented in
Limongi \& Chieffi (2003) and references therein. The only
difference in the setup of the code, compared to that discussed by
Limongi \& Chieffi (2003), deals with the mixing-length parameter
that now is set to $l/H_p$=1.5. The reason fors this choice is that we
wish to link the evolutionary models to the pulsation ones (see MFC)
that adopt such a value of $l/H_p$ to close the system of nonlinear
equations describing the dynamical structure of the envelope and the
convective flux (for details, see Stellingwerf 1982; Bono \&
Stellingwerf 1994; MFC).

\begin{table}[h]
\begin{center}
\begin{minipage}[t]{\columnwidth}
\caption{Selected parameters of the $Z$=0.0001 evolutionary tracks
without mass-loss. Mass ($M$) and luminosity ($L$) are in solar
units, while the evolutionary times $t$ are described in the
text. For the models which evolve into the IS, the last three
columns give the average luminosity and effective temperature at the
blue and the red edge of the IS (see text).} \label{massetempi}
\begin{tabular}{cccccccc}
\hline \hline
$M$ & $t$ & $M_{He}$ & $t_{He}$+$t_{EAGB}$ & $t_{IS}$ &$\langle$log$L_{IS}\rangle$ &$\langle$log$T_e$(FOBE)$\rangle$ &$\langle$log$T_e$(FRE)$\rangle$ \\
\hline
2.0 & 0.7 & 0.416 &  98+6 &  0.5 & 2.40 & 3.84 & 3.76 \\
1.8 & 1.0 & 0.425 & 104+7 &  3.2 & 2.30 & 3.84 & 3.76  \\
1.5 & 1.8 & 0.463 &  92+8 & 87.6 & 2.09 & 3.85 & 3.77  \\
1.3 & 2.8 & 0.483 &  76+9 & 35.0 & 2.03 & 3.85 & 3.77  \\
1.0 & 6.6 & 0.501 & 80+10   &-  &- &-  &- \\
0.8 & 14.5 & 0.508 & 79+11   & 69  & 1.77 & 3.86 & 3.78\\
\hline
\end{tabular}
\end{minipage}
\end{center}
\end{table}

For all the models, we give in Table 1 the evolutionary age ($t$, in
10$^9$ yr) and the He-core mass ($M_{He}$) at the beginning of the
central He-burning phase, the duration ($t_{He}$, in 10$^6$ yr) of
the central He-burning phase and the time ($t_{EAGB}$, in 10$^6$ yr)
elapsed from the central He-exhaustion and the beginning of the EAGB
phase.

Figure 1 shows the HR diagram of the models dealing with the central
He-burning phase. In order to point out the connection between
evolution and pulsation, we plot in this figure the difference between
the effective temperature of the model and the appropriate value at
the red edge of the fundamental pulsation region (FRE), which is also
the red limit of the whole instability strip. The pulsation models
computed by MFC show that for fixed metal content, mass and
luminosity, fundamental (F) pulsators are generally redder than first
overtone (FO) ones, so that the FRE and the blue edge of the first
overtone region (FOBE) can be taken as representative of the
boundaries of the whole instability strip. In this paper, the FRE is
computed according to Eq. (2a) in C04, while Eq. (1a) in C04 is used
to fix the FOBE at log$T_e$(FOBE)$-$log$T_e$(FRE)=0.08 (see last two
columns in Table 1). In such a way, the predicted pulsators of a given
mass are identified with the models whose effective temperature falls
between the FOBE and the FRE, i.e., with
log$T_e$(FOBE)$\ge$log$T_e\ge$log$T_e$(FRE).

Inspection of Fig. 1 shows that, in the absence of mass-loss,
He-burning models with $Z$=0.0001 and mass around $\sim$ 1-1.2\msun
evolve at effective temperatures lower than the FRE, yielding that no
pulsators are expected in this mass range. As for the models that
evolve into the IS, we list in Table 1 the lifetime ($t_{I}$, in
10$^6$ yr) and the time-averaged luminosity
($\langle$log$L_{IS}\rangle$) corresponding to the pulsation phase,
while the last two columns give the average effective temperature at
the blue (FOBE) and red (FRE) edges of the IS. The predicted pulsators
with $Z$=0.0001 and mass 1.3-2.0\msun follow a Mass-Luminosity ($ML$)
relation as $\langle$log$L\rangle \sim$ 1.77($\pm$0.05)+2.07log$M$, in
the absence of mass-loss.

\begin{figure}
\includegraphics[width=9cm]{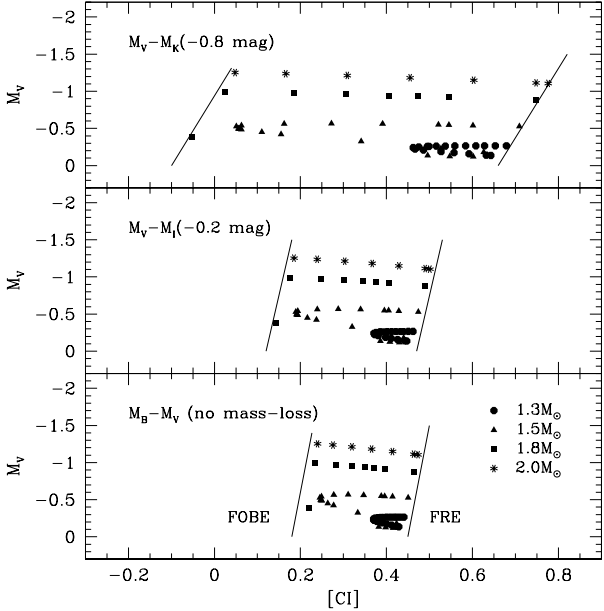}
\caption{\small{Selected Color-Magnitude diagrams of the predicted
pulsators with $Z$=0.0001 and the labeled masses, in absence of
mass-loss. The solid lines depict the blue (FOBE) and red (FRE)
boundaries of the pulsation region (see Table 2). As labeled,
the M$_V$-M$_K$ and M$_V$-M$_I$ colors are shifted by $-$0.8 mag
(top panel) and $-$0.2 mag (middle panel), respectively.}}
\end{figure}

Using the Castelli, Gratton \& Kurucz (1997 a,b) atmosphere models to
calculate the $BVRIJK$ magnitudes\footnote{$RI$ and $JK$ magnitudes
are in the Cousins (1980 and references therein) and Bessell \& Brett
(1988) photometric system, respectively.}, we show in Fig.  2 selected
Color-Magnitude ($CM$) diagrams of the predicted pulsators originating
from the evolutionary tracks presented in Fig.  1. The solid lines
depicting the blue (FOBE) and red (FRE) limits of the whole pulsation
region are listed in Table 2 together with the color uncertainty ($\pm
\epsilon$) which is due to the intrinsic uncertainty of the FOBE and
FRE effective temperatures (see MFC). Both
the FRE and the FOBE depend on the efficiency of convection in the
star external layers, namely on the adopted value of $l/H_p$ in the
pulsation model computations (Di Criscienzo et al. 2004 and references
therein, Fiorentino et al.  2006 in preparation). In particular,
when increasing the value of the mixing length parameter the FRE moves
towards higher effective temperatures whereas the FOBE has an
opposite behavior, at constant mass and luminosity. Since current
computations of the evolutionary models, based on the calibration of
the standard solar model, adopt $l/H_p \ge$ 1.5 (see Pietrinferni et
al. 2004) it follows that the lines drawn in Fig. 2 should represent
the bluest (FOBE) and reddest (FRE) limits of the AC instability strip
at $Z$=0.0001, in absence of the mass-loss.

\begin{table}[h]
\begin{center}
\begin{minipage}[t]{\columnwidth}
\caption{Predicted boundaries in the Color-Magnitude diagram of the
pulsator distribution generated by the evolutionary tracks in Fig.
1. For each given color $[CI]$, we give the value (mag) at the
absolute magnitude $M_V$=0 and $-$1.5 mag. In the last column, we
give the intrinsic uncertainty ($\pm \epsilon$) of the color.}
\label{colormagnitude}
\begin{tabular}{c|cc|cc|c}
\hline \hline
      &  FOBE        &  & FRE &   &       \\
 & $M_V$ & $M_V$   & $M_V$ & $M_V$ &$\pm \epsilon$ \\
\hline
$[CI]$      & 0     & $-$1.5  & 0     & $-$1.5 &  \\
\hline
$M_B-M_V$ & 0.18 & 0.23 & 0.45 & 0.50& 0.02\\
$M_V-M_R$ & 0.15 & 0.18 & 0.32 & 0.35& 0.02\\
$M_V-M_I$ & 0.32 & 0.38 & 0.67 & 0.73& 0.03\\
$M_V-M_J$ & 0.54 & 0.65 & 1.07 & 1.18& 0.04\\
$M_V-M_K$ & 0.70 & 0.86 & 1.46 & 1.62& 0.05\\
\hline
\end{tabular}
\end{minipage}
\end{center}
\end{table}

\begin{figure}
\includegraphics[width=9cm]{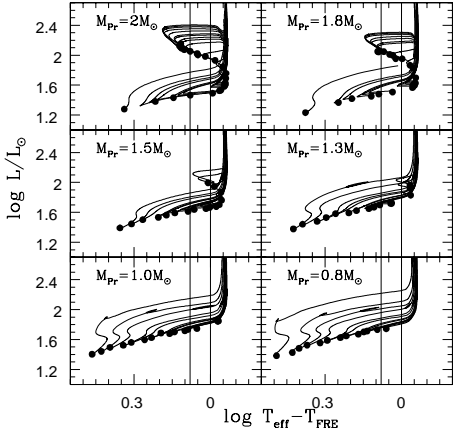}
\caption{\small{As in Fig. 1,
but for the ZAHB models generated from the labeled progenitor
mass.}}
\end{figure}

\begin{table}[h]
\begin{center}
\begin{minipage}[t]{\columnwidth}
\caption{ZAHB models for each progenitor star with
$Z$=0.0001 and mass $M_{pr}$. Models evolving from the post
ZAHB-turnover into the IS are in bold face, while those producing RR
Lyrae and Population II Cepheids are in italics. All the masses are in
solar units.} \label{modelli}
\begin{tabular}{c|ccccccc}
\hline \hline
$M_{pr}$&&&&$M$&&&\\
\hline
2.00&\bf{2.00}&\bf{1.98}&\bf{1.95}&\bf{1.90}&\bf{1.85}&\bf{1.78}&\bf{1.75}\\
&\bf{1.70}&\bf{1.65}&\bf{1.58}&\bf{1.38}&1.18&0.98&0.80\\
&0.78&0.76&0.74&0.72&0.70&0.62&0.60\\
&0.58&0.56&0.54&0.52&0.50\\
\hline
1.80&\bf{1.80}&\bf{1.78}&\bf{1.75}&\bf{1.70}&\bf{1.65}&\bf{1.58}&\bf{1.55}\\
&\bf{1.38}&1.18&0.98&0.80&0.78&0.76&0.74\\
&0.72&0.70&0.62&0.60&0.58&0.56&0.54\\
&0.50\\
\hline
1.50&\bf{1.50}&\bf{1.38}&\bf{1.18}&0.98&0.80&0.78&0.76\\
&0.74&0.72&0.70&0.68&0.66&0.64&0.62\\
&0.60&0.58&0.56\\
\hline
1.30&\bf{1.30}&\bf{1.18}&0.98&0.78&0.76&0.74&0.72\\
&0.68&0.66&0.64&0.62&0.60&0.58&0.56\\
&0.54\\
 \hline
1.00&1.00&0.98&0.78&0.76&0.74&0.72&0.70\\
&0.62&0.70&0.60&0.58&0.56\\
\hline
0.80&\it{0.80}&\it{0.78}&\it{0.76}&\it{0.74}&\it{0.72}&\it{0.70}&\it{0.62}\\
&\it{0.60}&\it{0.58}&\it{0.56}\\
\hline
\end{tabular}
\end{minipage}
\end{center}
\end{table}

\subsection{Models with mass loss}

The central He-burning models presented in Fig. 1, as characterized by
different values of $M_{He}$, do not properly define the so-called
Zero Age Horizontal Branch (ZAHB), which is the locus in the HR
diagram dealing with post He-flash stars that start to burn helium in
the center, having the same He-core mass but different total masses as
a consequence of mass loss during the Red Giant Branch phase.

In order to study the effects of mass loss, we have adopted progenitor
stars with the masses listed in Table 1 and for each progenitor mass
$M_{pr}$ we have computed a sequence of central He-burning models with
the same He-core mass of the progenitor [see column (3) in Table 1]
but with a total mass between $M_{pr}$ and a value slightly larger
than $M_{He}$. These models are listed in Table 3 and plotted in
Fig. 3 in the \lsun versus log$T_e-$FRE diagram. Inspection of this
figure, where each panel deals with a given progenitor star and the
vertical solid lines show the predicted boundaries of IS, clearly
shows that with $M_{pr}> $1.3\msun the effective temperature of the
ZAHB models decreases with increasing mass of the star, reaching a
minimum value corresponding to $\log{T_e}\sim$ 3.76 for models with
$\sim$ 1.0-1.2 $M_{\odot}$. After this minimum, both the luminosity
and the effective temperature of the more massive models start
increasing, forming an hook called the ``ZAHB turnover''. As a
consequence, only with $M_{pr}$ larger than $\sim$ 1.3\msun is there a
range of post ZAHB-turnover models that crossing the IS and that can
be identified with ACs. These models are reported in bold face in
Table 3, confirming that with $Z$=0.0001 the minimum mass for AC-like
pulsators is around 1.2\msun, also in the presence of mass
loss. Moreover, independently of the progenitor mass, the He-burning
evolution of models with mass from $\sim$ 1.0 to 1.2\msun proceeds at
an effective temperature lower than the FRE, whereas for those with
$M\le$ 0.8$M_{\odot}$ the crossing of the IS occurs at luminosity
levels that increase as the stellar mass decreases. Among the latter
models, those with $M_{pr}$=0.8\msun, as reported in italics in Table
3, generate the RR Lyrae stars and Population II Cepheids observed in
Galactic Globular Clusters (see C04 for details).

\begin{figure}
\includegraphics[width=8cm]{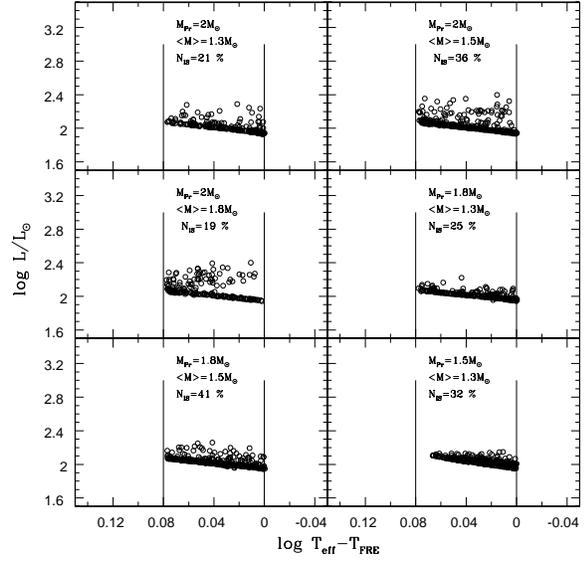}
\caption{\small{As in Fig. 1, but for the synthetic population of
pulsators (open circles) with a mean mass $\langle M\rangle$, as
produced by the labeled progenitor star with mass $M_{pr}$.}}
\end{figure}

\begin{figure}
\includegraphics[width=9cm]{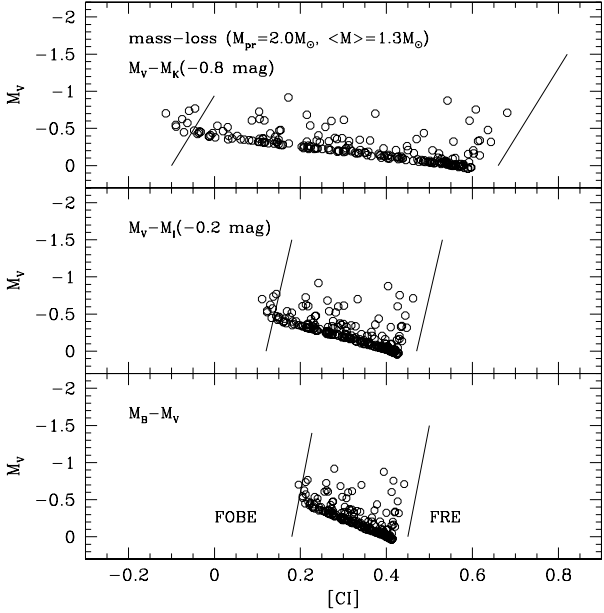}
\caption{\small{As in Fig. 2, but for the predicted pulsators with a
mean mass $\langle M\rangle$=1.3\msun, as generated by a 2.0\msun
progenitor star. The solid lines depict the boundaries in absence of
mass loss (see Table 2).}}
\end{figure}

\subsection{Synthetic populations}

The grid of models originating from a given progenitor mass is then
used to build up synthetic populations of AC pulsators.  In
particular, for each progenitor mass and its corresponding ZAHB
models, the synthetic population has been computed in the following
way:
\begin{enumerate}

\item We fix a mass Gaussian distribution centered on a chosen mean
mass $\langle M\rangle$ and with $\sigma$=0.2\msun in order to account
for some spread in the amount of mass loss;

\item We randomly extract a mass value weighted over this Gaussian
distribution;

\item For each extracted mass, we randomly extract an evolutionary
lifetime in the range between the central He-burning and the EAGB
phase;

\item The extracted mass is then located in the HR diagram according
to the extracted time;

\item For each extracted mass and time we evaluate the luminosity and
effective temperature;

\item The number of extractions is fixed to 1000 in order to have
good statistics.
\end{enumerate}

In this framework, each simulation is characterized by a progenitor
mass $M_{pr}$ and a mean mass $\langle M\rangle$, as shown in Fig.  4,
where the predicted pulsators obtained for three progenitor stars
(2.0, 1.8, 1.5\msun) and different choices of $\langle M\rangle$ are
presented. In each panel of this figure, in addition to the progenitor
mass and the mean mass of the pulsators, we give also the percentage
($N_{IS}$) of the central He-burning stars populating the IS.

All the pulsators plotted in a given panel of Fig. 4 have the same
$M_{He}$ of the progenitor star and that in the explored mass range
the He-core mass decreases as $M_{pr}$ increases (see Table 1). Since,
everything else being constant, the star luminosity increases with
$M_{He}$, it follows that the pulsators with a given mass originating
in the presence of mass loss are fainter than those in absence of mass
loss. At constant metal content and total mass, the decrease in
luminosity follows the variation of the He-core mass: as an example,
with $Z$=0.0001 the 1.3\msun pulsators in absence of mass loss
($M_{He}$=0.483\msun) have an average visual magnitude $M_V \sim
-$0.24 mag, while those generated by a progenitor mass 1.5\msun
($M_{He}$=0.463\msun), 1.8 ($M_{He}$=0.425\msun) and 2.0\msun
($M_{He}$=0.416\msun) have $M_V \sim -$0.17, $-$0.07 and $-$0.03 mag,
respectively.

As shown by the pulsation models computed by MFC, a luminosity
decrease at constant mass causes the entire pulsation region to move
towards larger effective temperatures, such that the resulting $CM$
distribution of the predicted pulsators becomes bluer ($\sim$ 0.03 mag
with $B-V$ and $\sim$ 0.07 mag with $V-K$) than in the absence of mass
loss. This is shown in Fig.  5, where selected $CM$ distributions of
the pulsators with $\langle M\rangle$=1.3\msun, as generated by
$M_{pr}$=2.0\msun, are compared with the boundaries in the absence of
mass loss (solid lines). Thus, the predicted FRE values given in Table
2 can be taken as the reddest limits of the AC instability strip at
$Z$=0.0001 also in presence of a mass loss of up to $\sim$
0.7\msun.

\begin{figure}
\includegraphics[width=8cm]{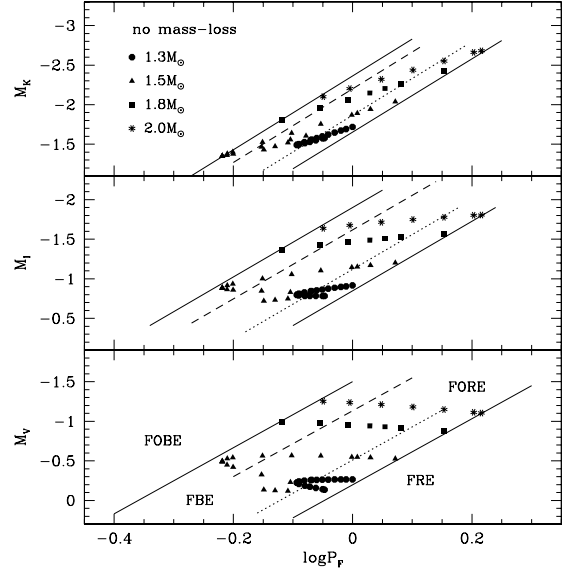}
\caption{\small{Selected Period-Magnitude diagrams of the predicted
pulsators with $Z$=0.0001 and the labeled mass, in the absence of mass
loss. The solid lines depict the blue (FOBE) and red (FRE) limits of
the entire pulsator distribution, while the dashed and dotted lines
are the blue edge for the fundamental (FBE) and the red edge for first
overtone (FORE) pulsation, respectively. All the pulsators
are plotted with their fundamental period, irrespective of the actual
limits for the fundamental or first overtone pulsation mode (see
text).}}
\end{figure}

\section{Pulsational relations}
\subsection{Period-Magnitude and Period-Color relations}

In analogy with other radial pulsators such as RR Lyrae stars and
Classical Cepheids, the pulsation period of ACs is uniquely defined
by the intrinsic stellar parameters: mass, luminosity, and effective
temperature. Based on the pulsation models presented by MFC and C04,
for the fundamental (F) mode we adopt
$$\log P_F=10.88+0.82\log L-0.62\log M-3.31\log T_e,\eqno(1)$$
\noindent where the mass $M$ and the luminosity $L$ are in solar
units. As for the first overtone (FO) period, from the MFC models we
estimate that the average difference between the computed periods and
the values obtained from equation (1) is log$P_{FO}$=log$P_F-$0.13, at
constant mass, luminosity and effective temperature.

\begin{table}[h]
\begin{center}
\caption{Predicted boundaries in the Period-Magnitude diagram of the
pulsator distribution with $Z$=0.0001 and $l/H_p$=1.5, in the absence
of mass loss. The red (FRE) and blue (FBE) limits for fundamental
pulsators are given as a function of the fundamental period $P_F$,
while those (FORE and FOBE) for FO pulsators as a function of the
first overtone period $P_{FO}$ (see text). } \label{massetempi}
\begin{tabular}{c|cc|c}
\hline \hline
\multicolumn{4}{c}{$M_i=a+b$log$P_F$}\\
\hline
&FRE&&FBE\\
\hline
$M_i$ &$b$     &  $a$              &$a$\\
\hline
$M_V$ &$-$4.17 &  $-$0.20$\pm$0.08 & $-$1.13$\pm$0.11   \\
$M_R$ &$-$4.27 &  $-$0.50$\pm$0.08 & $-$1.35$\pm$0.11   \\
$M_I$ &$-$4.37 &  $-$0.85$\pm$0.07 & $-$1.62$\pm$0.10   \\
$M_J$ &$-$4.50 &  $-$1.30$\pm$0.06 & $-$1.93$\pm$0.08   \\
$M_K$ &$-$4.65 &  $-$1.65$\pm$0.05 & $-$2.20$\pm$0.07  \\
\hline
\hline
\multicolumn{4}{c}{$M_i=a+b$log$P_{FO}$}\\
\hline
      &FORE&&FOBE\\
\hline
$M_i$ & $b$    & $a$              &$a$ \\
\hline
$M_V$ &$-$4.17 & $-$1.05$\pm$ 0.13 &$-$2.04$\pm$0.08   \\
$M_R$ &$-$4.27 & $-$1.35$\pm$ 0.13 &$-$2.26$\pm$0.08   \\
$M_I$ &$-$4.37 & $-$1.69$\pm$ 0.11 &$-$2.47$\pm$0.07   \\
$M_J$ &$-$4.50 & $-$2.10$\pm$ 0.10 &$-$2.69$\pm$0.06   \\
$M_K$ &$-$4.65 & $-$2.47$\pm$ 0.08 &$-$2.95$\pm$0.05   \\
\hline
\end{tabular}
\end{center}
\end{table}

Adopting the fundamental period given by Eq. (1), we show in Fig. 6
selected Period-Magnitude ($PM$) diagrams for all the predicted
pulsators in the absence of mass loss, as presented in Fig.  2. In
this figure, the solid lines showing the limits of the pulsator
distribution are the boundaries (FOBE and FRE) of the whole
instability strip, adopting for all the pulsators the period
corresponding to the fundamental mode. In reality, the MFC models show
that within the instability strip there exist a blue edge for
fundamental (FBE) and a red edge for first overtone pulsation (FORE),
as given by log$T_e$(FBE)$-$log$T_e$(FRE)=0.055$\pm$0.005 and
log$T_e$(FORE)$-$log$T_e$(FRE)=0.018$\pm$0.008, for fixed mass and
luminosity. For effective temperatures between the FORE (FOBE) and the
FRE (FBE) only the F (FO) mode is efficient, whereas between the FBE
(dashed line in Fig. 6) and the FORE (dotted line in Fig. 6) both the
pulsation modes are possible. However, if the first overtone period is
adopted for FO pulsators, then both the FOBE and the FORE should be
shifted by $\delta$log$P=-$0.13, with the result that the FORE and the
FBE become almost coincident.  This is shown in Table 4, where the
limits for the fundamental pulsator distribution (FBE and FRE) are
given as a function of $P_F$, while those for first overtone pulsators
(FOBE and FORE) as a function of $P_{FO}$.

By inspection of Table 4 and Fig. 6, one has that the range in
magnitudes, at a fixed period, expected for the predicted F or FO
pulsators is quite large in the optical bands, decreasing when moving
to longer wavelengths. Thus, synthetic $PM_V$ or $PM_I$ relations will
significantly depend on the distribution of the pulsators within the
pulsation region, at variance with the case of near-infrared
magnitudes. In particular, the mean relations
$$M^F_K=-1.93(\pm 0.15)-4.65\log P_F,\eqno(2a)$$
\noindent defined by all the F pulsators falling between the FBE and the FRE, and
$$M^{FO}_K=-2.71(\pm 0.18)-4.65\log P_{FO},\eqno(2b)$$
\noindent by all the FO pulsators between the FOBE and the FORE, can
be used for distance determinations. Moreover, since in the
$M_K$-log$P$ plane the dispersion at fixed periods depends almost
entirely on the pulsator mass, we {\bf derive} that the
least-squares solution to all the $Z$=0.0001 pulsators, taken with
their fundamental period, yields a quite tight mass-dependent $PM_K$
relation, as given by
$$M_K=-1.40(\pm 0.04)-2.44\log P_F-2.57\log M,\eqno(3)$$

\begin{figure}
\includegraphics[width=8cm]{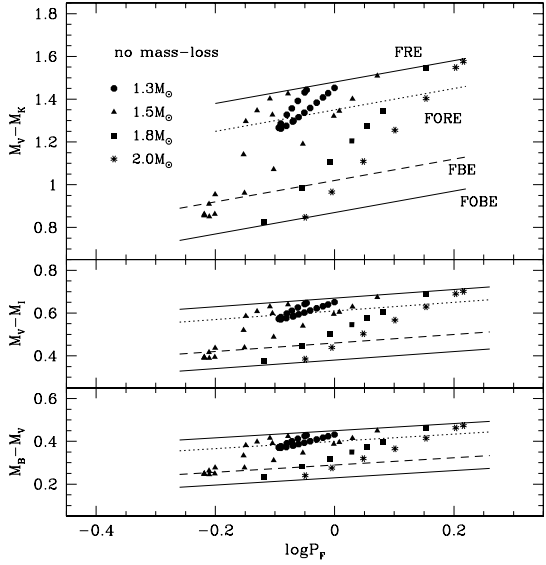}
\caption{\small{Selected Period-Color diagrams of the pulsators with
$Z$=0.0001 and the labeled mass, in the absence of mass loss. The
solid lines depict the limits (FOBE and FRE) of the entire instability
strip, while the dashed and the dotted lines are the blue edge for
fundamental (FBE) and the red edge for first overtone (FORE)
pulsation, respectively. Note that all the pulsators and the pulsation
edges are plotted with the fundamental period.}}
\end{figure}

\begin{table}[h]
\begin{center}
\caption{As in Table 4, but for the Period-Color diagrams.}
\label{massetempi}
\begin{tabular}{c|cc|c}
\hline \hline
\multicolumn{4}{c}{$[CI]=a+b$log$P_F$}\\
\hline
        &FRE&&FBE\\
\hline
$[CI]$ & $b$ & $a$ & $a$ \\
\hline
$M_B-M_V$ &0.17 &0.45$\pm$0.03 &  0.29$\pm$0.04 \\
$M_V-M_R$ &0.10 &0.32$\pm$0.03 &  0.21$\pm$0.04  \\
$M_V-M_I$ &0.20 &0.67$\pm$0.04 &  0.46$\pm$0.06  \\
$M_V-M_J$ &0.35 &1.09$\pm$0.05 &  0.76$\pm$0.07  \\
$M_V-M_K$ &0.50 &1.48$\pm$0.06 &  1.02$\pm$0.08  \\
\hline
\hline
\multicolumn{4}{c}{$[CI]=a+b$log$P_{FO}$}\\
\hline
&FORE&&FOBE\\
\hline
$[CI]$ & $b$ & $a$ & $a$ \\
\hline
$M_B-M_V$ &0.17 &0.42$\pm$0.05 & 0.25$\pm$0.03 \\
$M_V-M_R$ &0.10 &0.29$\pm$0.05 & 0.18$\pm$0.03 \\
$M_V-M_I$ &0.20 &0.64$\pm$0.06 & 0.41$\pm$0.04 \\
$M_V-M_J$ &0.35 &1.04$\pm$0.08 & 0.70$\pm$0.05 \\
$M_V-M_K$ &0.50 &1.42$\pm$0.10 & 0.94$\pm$0.06 \\
\hline
\end{tabular}
\end{center}
\end{table}

For the Period-Color ($PC$) diagram, Fig. 7 shows that the dispersion
of the pulsator distribution, at fixed period, increases moving from
$B-V$ to $V-K$ colors. As a consequence we find that only the
mean optical relations derived from all the predicted F pulsators

$$[M_B-M_V]^F=0.37(\pm 0.05)+0.17\log P_F\eqno(4a)$$
$$[M_V-M_I]^F=0.57(\pm 0.07)+0.20\log P_F\eqno(4b)$$
\noindent and the ones derived for FO pulsators
$$[M_B-M_V]^{FO}=0.34(\pm 0.07)+0.17\log P_{FO}\eqno(4c)$$
$$[M_V-M_I]^{FO}=0.52(\pm 0.09)+0.20\log P_{FO}\eqno(4d)$$
\noindent

appear useful for reddening determinations.
We notice that in this case, as reported in Table 5, the adoption of
the first overtone period for FO pulsators will leave a FORE
significantly redder than the blue edge of F pulsators, taken with
their fundamental period.

\subsection{Period-Magnitude-Color relations}
The application of Eq. (1) in the observational plane is a
mass-dependent Period-Magnitude-Color relation (hereafter, named
$PMC$) in which the pulsator absolute magnitude is strictly correlated
with period and color, for any given mass. A linear regression through
the magnitudes and the fundamental periods of all the predicted
pulsators generated in the absence of mass loss yields the
mass-dependent $PMC$ relations listed in Table 6.

\begin{table}[h]
\begin{center}
\caption{Mass-dependent $PMC$ relations. In
the last column, we give the total intrinsic dispersion $\pm \sigma_V$.}
\label{PLCM}
\begin{tabular}{ccccccc}
\hline
\hline
\multicolumn{6}{c}{$M_V=a+b$log$P_F+c[CI]+d$log$M$}\\
\hline
$[CI]$ & $a$ & $b$ & $c$ & $d$ &$\sigma_V$ \\
\hline
$M_B-M_V$ & $-$1.75 & $-$2.86 & 3.99 & $-$2.02 & 0.03 \\
$M_V-M_R$ & $-$1.86 & $-$2.99 & 5.92 & $-$1.88 & 0.03 \\
$M_V-M_I$ & $-$1.99 & $-$3.02 & 2.96 & $-$1.85 & 0.03 \\
$M_V-M_J$ & $-$2.01 & $-$3.03 & 1.93 & $-$1.88 & 0.03 \\
$M_V-M_K$ & $-$1.93 & $-$3.02 & 1.29 & $-$1.90 & 0.03 \\
\hline
\end{tabular}
\end{center}
\end{table}

\begin{figure}
\includegraphics[width=8cm]{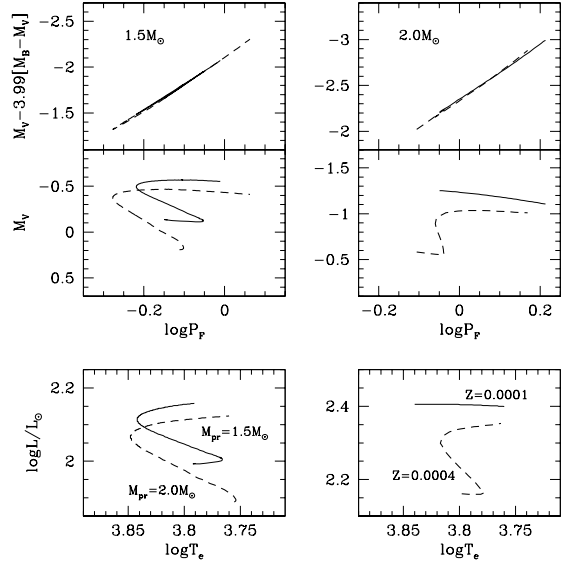}
\caption{\small{{\it Left panels} - Behavior of $Z$=0.0001 pulsators
with $M$=1.5\msun as generated in the absence of mass loss (solid line)
or by a progenitor with 2.0\msun (dashed line). In the lower two panels,
we plot the luminosity versus the effective temperature
and the visual magnitude versus the fundamental period,
while the upper panel deals with the $PMC$ relation given in Table
6. {\it Right panels} - As in the left panels, but with 2.0\msun
pulsators with $Z$=0.0001 (solid line) and $Z$=0.0004 (dashed line),
in the absence of mass loss.}}
\end{figure}

The mass-dependent relations hold for all the combinations of mass and
luminosity, thus they are not influenced by the occurrence of mass
loss before or during the He-burning phase.  As an example, the left
panels in Fig. 8 show the comparison between the 1.5\msun model
without mass loss ($M_{He}$=0.463\msun, solid line) and the 1.5\msun
structure generated by a 2.0\msun progenitor ($M_{He}$=0.416\msun,
dashed line) which lost 0.5\msun.  According to the different He-core
masses, the two models evolve at different bolometric luminosities and
they follow distinct behaviors in the $M_V$-log$P_F$ plane (lower two
panels), with an average visual magnitude $M_V\sim -$0.41 mag and
$\sim -$0.28 mag without and with mass loss, respectively. However,
the effects of such a luminosity variation on the $PMC$ relation
(upper panel) are zero. Similar results are derived when the variation
in luminosity is due to a different metal content, at least in the
range $Z$=0.0001-0.0004. This is shown in the right panels in Fig. 8,
where the comparison between 2.0\msun models with $Z$=0.0001 (solid
line) and $Z$=0.0004 (dashed line) is presented. Later, we will give a
full discussion of the metallicity effects on the various pulsational
relations.

The constraints provided by the evolution theory, and in particular
the occurrence of a $ML$ relation that binds the permitted values of
mass and luminosity of the predicted pulsators, allow us to derive
``evolutionary'' $PMC$ relations (hereafter $PMC_e$) where the
mass-term is removed. Obviously these relations are expected to depend
on the assumption on the amount of mass loss.

\begin{table}[h]
\begin{center}
\caption{Evolutionary $PMC_e$ relations for $Z$=0.0001 pulsators in
  the mass range 1.3-2.0\msun, in the absence of mass loss.}
\label{PLCM}
\begin{tabular}{cccc}
\hline
\hline
\multicolumn{4}{c}{$M_V=\alpha_0+\beta_0$log$P_F+\gamma_0[CI]$}\\
\hline
[CI] & $\alpha_0$ & $\beta_0$ & $\gamma_0$ \\
\hline
$M_B-M_V$ & $-$2.54$\pm$0.08 & $-$4.05$\pm$0.06 & 4.91$\pm$0.10 \\
$M_V-M_R$ & $-$2.71$\pm$0.08 & $-$4.18$\pm$0.06 & 7.58$\pm$0.16 \\
$M_V-M_I$ & $-$2.89$\pm$0.08 & $-$4.21$\pm$0.06 & 3.82$\pm$0.08 \\
$M_V-M_J$ & $-$3.02$\pm$0.08 & $-$4.30$\pm$0.06 & 2.48$\pm$0.05 \\
$M_V-M_K$ & $-$3.00$\pm$0.08 & $-$4.37$\pm$0.06 & 1.82$\pm$0.03 \\
\hline
\end{tabular}
\end{center}
\end{table}

For the sample of $Z$=0.0001 pulsators with mass 1.3, 1.5, 1.8 and
2.0\msun in the absence of mass loss, i.e., in the case of a mass
variation due to different evolutionary ages (see $t$ values in Table
1), a linear regression through all the predicted pulsators, taken
with their fundamental period, yields the $PMC_e$ relations given in
Table 7. These relations can be used to determine the distance to
individual Cepheids with a formal accuracy of $\sim$ 0.1 mag, provided
that the intrinsic colors are known with the adequate accuracy.

\begin{figure}
\includegraphics[width=8cm]{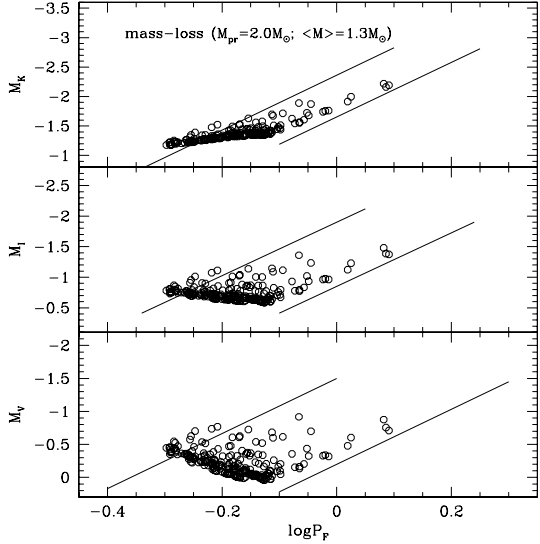}
\caption{\small{As in Fig. 6, but for the predicted pulsators with a mean mass
$\langle M\rangle$=1.3\msun, as generated by a 2.0\msun progenitor
star. The solid lines depict the limits (FOBE and FRE) of the whole instability strip
in absence of mass loss.}}
\end{figure}

\begin{figure}
\includegraphics[width=8cm]{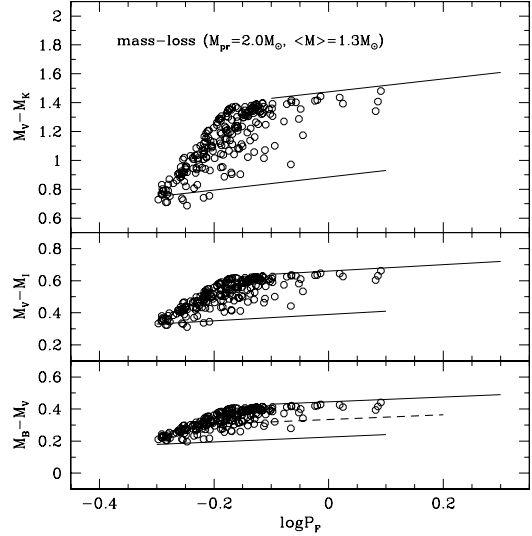}
\caption{\small{As in Fig. 7, but for the predicted pulsators with a
mean mass $\langle M\rangle$=1.3\msun, as generated by a 2.0\msun
progenitor star. The solid lines show the limits (FOBE and FRE) of the
IS in the absence of mass loss.}}
\end{figure}

For coeval pulsators whose mass variation is due to different amounts
of mass loss suffered by the same progenitor star, we show in Fig. 9
and Fig. 10 the $PM$ and $PC$ distributions of the synthetic pulsators
with a mean mass $\langle M\rangle$=1.3\msun, as generated by a mass
loss of $\sim$ 0.7\msun in a 2.0\msun progenitor star. For a given
mass, these pulsators are fainter than those originating in the
absence of mass loss and that following such a luminosity decrease the
pulsation edges move towards larger effective temperatures, i.e.,
shorter periods.  Eventually, these two effects yield that the whole
$PM$ distribution is on average brighter by about 0.1 mag than that in
absence of mass loss (solid lines), for a fixed metal content.
Concerning the $PC$ distribution, the concomitant variation of color
and period causes the limits of the instability strip to be not
significantly different from those in the absence of mass loss (solid
lines).

\begin{table}[h]
\begin{center}
\caption{Evolutionary $PMC_e$ relations for $Z$=0.0001 pulsators
generated by mass loss in a 2.0\msun progenitor star.}
\label{PLCM}
\begin{tabular}{cccc}
\hline
\hline
\multicolumn{4}{c}{$M_V=\alpha+\beta$log$P+\gamma[CI]$}\\
\hline
[CI] & $\alpha$ & $\beta$ & $\gamma$ \\
\hline
$M_B-M_V$ & $-$2.36$\pm$0.06 & $-$3.15$\pm$0.06 & 4.61$\pm$0.04\\
$M_V-M_R$ & $-$2.55$\pm$0.06 & $-$3.45$\pm$0.06 & 7.12$\pm$0.06 \\
$M_V-M_I$ & $-$2.71$\pm$0.06 & $-$3.52$\pm$0.06 & 3.58$\pm$0.03\\
$M_V-M_J$ & $-$2.87$\pm$0.06 & $-$3.68$\pm$0.06 & 2.34$\pm$0.02 \\
$M_V-M_K$ & $-$2.86$\pm$0.06 & $-$3.79$\pm$0.06 & 1.72$\pm$0.01 \\
\hline
\end{tabular}
\end{center}
\end{table}

For the $PMC_e$ relations in the presence of mass loss, the linear
regression through the various synthetic populations presented in
Fig. 4 shows that the coefficients are slightly dependent on the mass
of the progenitor star. Using all the predicted pulsators generated by
a 2.0\msun progenitor, we find the coefficients listed in Table 8.
Moreover, we find that these relations hold also with
$M_{pr}$=1.5\msun and 1.8\msun, with only the constant term depending
on the mass, or age, of the progenitor star as $\delta
\alpha$=$-$0.16($M_{pr}-$2.0).

\begin{figure}
\includegraphics[width=8cm]{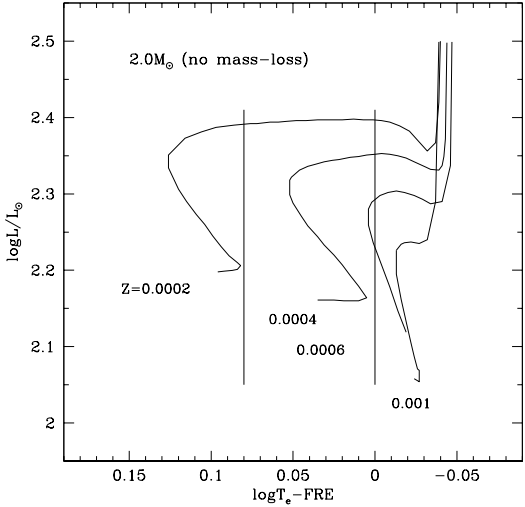}
\caption{\small{As in Fig. 1, but for 2.0\msun evolutionary tracks
with the labeled metal content, in absence of mass loss.}}
\end{figure}

\begin{figure}
\includegraphics[width=8cm]{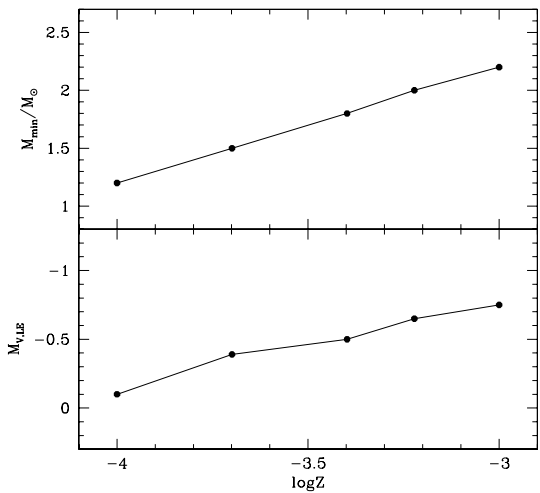}
\caption{\small{{\it (upper panel)} - Minimum mass for the
occurrence of massive central He-burning pulsators as a function of
the metal content. {\it (lower panel)} - Predicted faintest
magnitude ($M_{V,LE}$) of the massive pulsators as a function of
metal content.}}
\end{figure}

\subsection{Metallicity effects and mean magnitudes}

In order to discuss the effects of the metal content on the various
pulsational relations, we first present some basic evolutionary
constraints. This issue has been discussed in previous papers (see C04
and references therein); in Fig. 11 the behavior of selected
evolutionary tracks (Cariulo, Degl'Innocenti \& Castellani
2004)\footnote{See also http://gipsy.cjb.net in the ``Pisa
Evolutionary Library''} with $M$=2.0\msun and metal content varying
from $Z$=0.0002 to $Z$=0.001, are shown to highlight that the blueward
extension of the central He-burning path becomes fainter and redder
with increasing the metal content, at a given mass. As a consequence,
the minimum mass evolving from the ZAHB turnover into the IS, namely
the minimum mass for the occurrence of massive central He-burning
pulsators, increases with increasing $Z$, passing from $\sim$ 1.2\msun
at $Z$=0.0001 to $\sim$ 1.7\msun with $Z$=0.0004 and to $\sim$
2.2\msun with $Z$=0.001 (see upper panel in Fig. 12). Correspondingly,
the lower luminosity level ($M_{V,LE}$) of the massive pulsators
increases from $\sim -$0.1 mag ($Z$=0.0001) to $\sim -$0.5 mag at
$Z$=0.0004 and $\sim -$0.8 mag at $Z$=0.001 (see lower panel in
Fig. 12 as well as Fig. 7 in C04). Thus, the intrinsically faint ACs
observed around $\sim$ 0 mag (see following Section 4) imply that the
metal content cannot be larger than $Z\sim$ 0.0004.

An increase of the metal content from $Z$=0.0001 to $Z$=0.0004 reveals
that the average luminosity of the predicted AC pulsators with a given
mass decreases as $\Delta$log$L\sim -$0.1. The effects of such a
luminosity variation on the boundaries of the IS are similar to those
discussed in the case of mass loss: the $PM$ distribution, and the
limits given in Table 4, become brighter by about 0.1 mag, while the
effects on the $PC$ distribution are negligible. Concerning the
evolutionary $PMC_e$ relations, we derive that the correction to the
$Z$=0.0001 zero-points ($\alpha_0$ in Table 7 and $\alpha$ in Table 8)
are $\sim -$0.10 mag.

The magnitudes computed so far are the static values the stars would
have if they were not pulsating, whereas the measurements deal with
time-averaged quantities over a pulsation period. The mean values may
be significantly different from the static ones (see , e.g., MFC;
Marconi et al. 2003; Di Criscienzo, Marconi \& Caputo 2004) and thus
the coefficients of the mass-dependent $PMC$ relations we give in
Table 6 are slightly different from those derived by MFC on the basis
of intensity-averaged magnitudes of the pulsation models (see their
Table 3). The mass-dependent $PMC$ relations provided by MFC, as based
on intensity-weighted magnitudes, should be preferred to estimate the
mass of individual variables with well-measured absolute magnitudes
and intrinsic colors; alternatively, if the intrinsic color and the
mass are known, they give accurate individual distances.

For near-infrared bands the discrepancy between static and mean values
is always insignificant.  In summary, the boundaries given in Table 5
and Table 6 and the mean relations derived in Section 3.1 can be
safely compared with measured intensity-weighted magnitudes, whereas
the effects on the $PMC_e$ relations dealing with optical colors may
reach significant values. As an example, the quantity $\langle
V\rangle -\gamma_0[\langle B\rangle-\langle V\rangle$] may be up to
$\sim$ 0.14 mag fainter than the  corresponding static value.

\begin{figure}
\includegraphics[width=8cm]{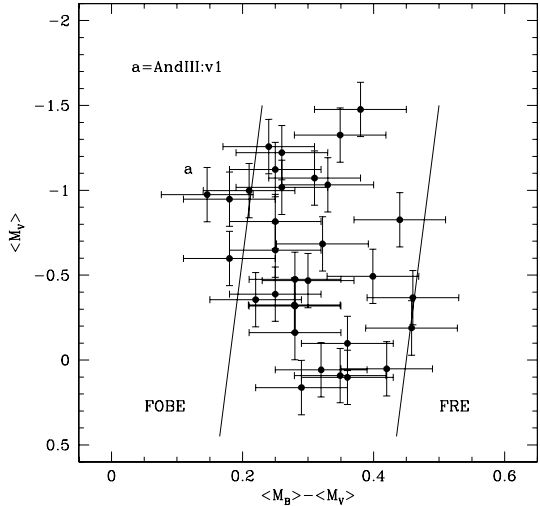}
\caption{\small{Color-Magnitude diagram of observed ACs (filled
    circles) compared to the predicted boundaries given in Table 2
    (solid lines).}}
\end{figure}

\begin{figure}
\includegraphics[width=8cm]{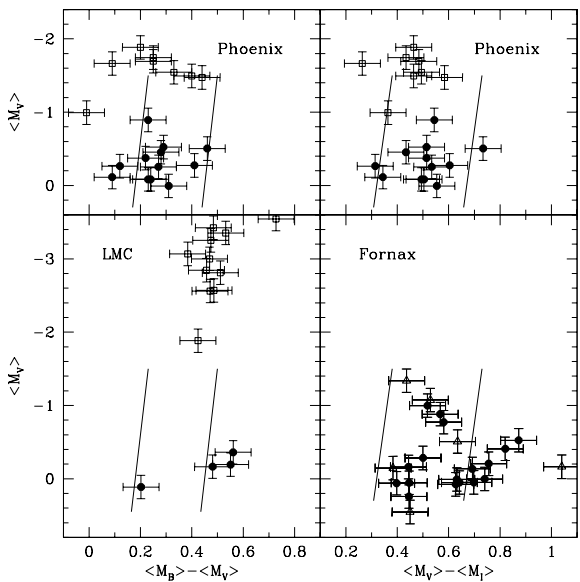}
\caption{\small{Color-Magnitude diagram of observed ACs (filled
circles) compared to the predicted boundaries given in Table 2.  We
plot also short-period Classical Cepheids (spCC: open squares) and
Population II Cepheids (P2C: open triangles).}}
\end{figure}

\section{Comparison with observations}

For the dwarf galaxies listed in Table 9, Figs. 13 and 14 show the
intensity-weighted absolute magnitude\footnote{In this paper, we adopt
a typical extinction law (Cardelli et al. 1989) with $A_V$=3.1$E(B-V)$
and $E(V-I)$=1.3$E(B-V)$.} of the observed ACs (filled circles) versus
the intrinsic color together with the predicted limits of the
instability strip (solid lines), as given in Table 2. Although these
limits hold with $Z$=0.0001 in the absence of mass loss, we have
already shown that even with a mass loss of $\sim$ 0.7\msun the color
shift is only $\sim -$0.03 mag (see Fig. 5) and that the metallicity
effect is negligible.  The errorbars are drawn adopting a photometric
error of $\pm$0.05 mag on each mean magnitude and an uncertainty of
0.15 mag on the intrinsic distance modulus. This yields $\epsilon
(M_V)=\pm$0.16 mag and $\epsilon (CI)=\pm$0.07 mag.

\begin{table}[h]
\begin{center}
\caption{Dwarf galaxies with observed Anomalous Cepheids listed with
their intrinsic distance modulus $\mu_0$, $E(B-V)$ reddening, and
metal abundance [Fe/H].}
\label{PLCM}
\begin{tabular}{lccccc}
\hline\hline
galaxy  &$\mu_0$& $E{B-V}$ & [Fe/H] &  Ref. & Notes \\
        &  (mag)&  (mag)    &        &       & \\
\hline
AndI    & 24.5 & 0.05 & $-$1.5 &  1 &  a   \\
AndII   & 24.1 & 0.06 & $-$1.5 &  2 &  a \\
AndIII  & 24.3 & 0.06 & $-$1.9 &  1 &  a  \\
AndVI   & 24.5 & 0.06 & $-$1.6 &  3 &  a \\
Carina  & 20.1 & 0.04 & $-$1.6 &  4 &  a \\
Draco   & 19.5 & 0.03 & $-$2.1 &  3 &  a  \\
Fornax  & 20.7 & 0.03 & $-$1.6 &  5 &  b,e    \\
LeoII   & 21.6 & 0.02 & $-$1.9 &  3,6& c \\
LMC     & 18.5 & 0.10 & $-$1.7 & 7   & a,f   \\
Phoenix & 23.1 & 0.02 & $-$1.4 & 8   & d,f   \\
Sculptor& 19.6 & 0.02 & $-$1.8 & 9,10& c \\
Sextans & 19.7 & 0.04 & $-$1.6 & 11  & a\\
\hline
\end{tabular}
\end{center}
{\it References} - (1): Pritzl et al. 2005; (2): Pritzl et al. 2004;
(3): Pritzl et al. 2002; (4): Dall'Ora et al. 2003; (5): Bersier \&
Wood 2002; (6): Siegel \& Majewski 2000; (7): Di Fabrizio et al.
2005; (8): Gallart et al. 2004; (9): Kaluzny et al. 1995; (10):
Clementini et al. 2005; (11): Mateo, Fischer \& Krzeminski 1995.
{\it Notes} - (a): $B,V$ data; (b): $V,I$ data; (c): $V$ data; (d):
$B,V,I$ data; (e) Population II Cepheids also observed; (f):
Short-period Classical Cepheids also observed.
\end{table}

In Fig. 13, we find an agreement between the observed $CM$
distribution and the predicted limits (FOBE and FRE) of the
instability strip, with the only exception of v1 in AndIII whose color
is significantly bluer than the predicted FOBE. As for the galaxies in
Fig. 14, the agreement is less because of the several ACs (filled
circles) located out of the predicted IS (see column (2) in Table
10). However, in this diagram the observed Population II Cepheids
(P2C: open triangles) are not clearly separated from ACs, while the
short-period Classical Cepheids (spCC: open squares) agree with the AC
boundaries extended to brighter magnitudes.

\begin{figure}
\includegraphics[width=8cm]{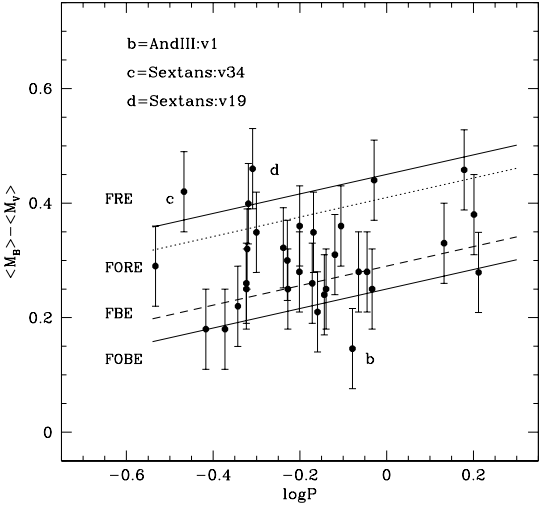}
\caption{\small{Period-Color diagram of observed ACs (filled circles)
    compared to the predicted boundaries for F and FO pulsators, as
    given in Table 5.}}
\end{figure}

\begin{figure}
\includegraphics[width=8cm]{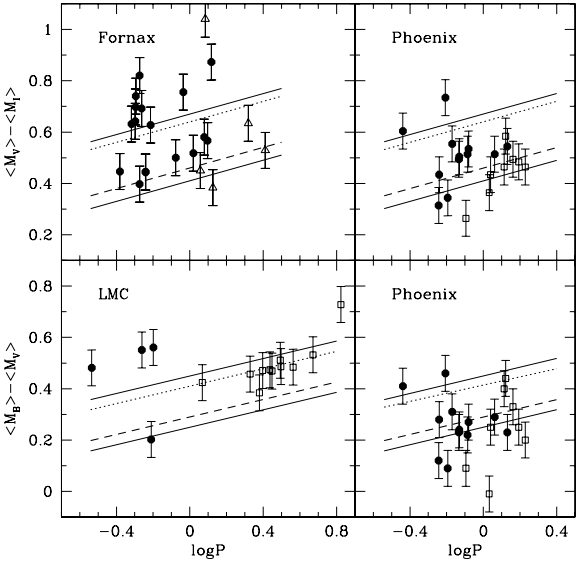}
\caption{\small{Period-Color diagram of observed ACs (filled circles)
compared to the predicted boundaries for F and FO pulsators, as
given in Table 5.  We plot also short-period Classical Cepheids (spCC:
open squares) and Population II Cepheids (P2C: open triangles).}}
\end{figure}

For the $PC$ diagrams, Figs. 15 and 16 show the intrinsic colors
versus the observed periods in compared to the predicted limits.

For the galaxies plotted in Fig. 15, we find an agreement between the
observed distribution and the predicted FOBE and FRE, with the
exception of variable v1 in AndIII, which is bluer than the FOBE, and
v19 and v34 in Sextans, which appear somehow redder than the FRE.
Concerning the galaxies in Fig. 16, we have again a significant number
of variables lying out of the predicted IS, as listed in column (3) of
Table 10.

\begin{figure}
\includegraphics[width=8cm]{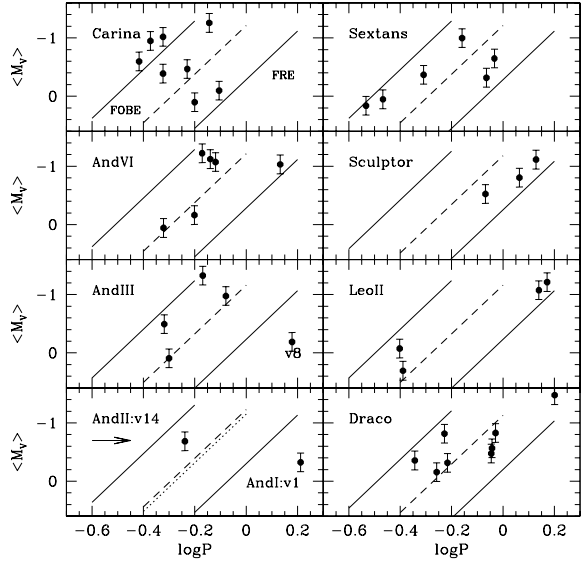}
\caption{\small{Period-Magnitude diagram of observed ACs (filled
circles) in comparison with the predicted boundaries
for F and FO pulsators, as given in Table 4, corrected for metallicity.}}
\end{figure}

\begin{figure}
\includegraphics[width=8cm]{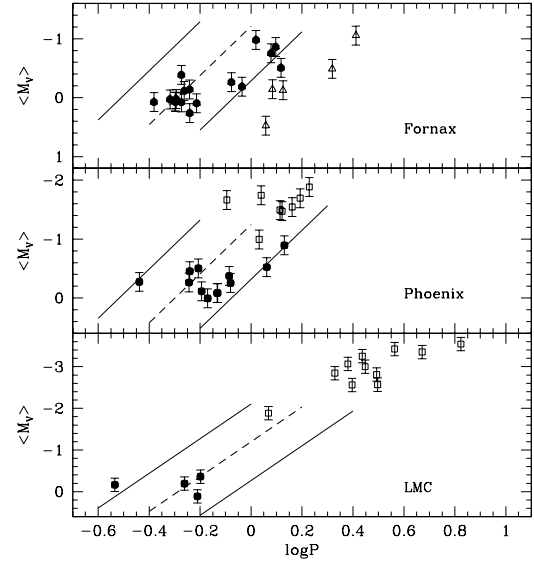}
\caption{\small{Period-Magnitude diagram of observed ACs (filled
circles) in comparison with the predicted boundaries
for F and FO pulsators, as given in Table 4, corrected for metallicity.
We plot also short-period Classical Cepheids (spCC: open
squares) and Population II Cepheids (P2C: open triangles). The solid
lines are the predicted boundaries given in Table 4 corrected for
metallicity.}}
\end{figure}

Figs. 17 and 18 show the absolute magnitudes versus the observed
period. In this case, since the zero-points listed in Table 4 hold in
the absence of mass loss and with $Z$=0.0001, i.e., at [Fe/H]=$-$2.1
referenced to $Z_{\odot}$=0.012 (Asplund et al. 2005), they are
corrected for the metal content of the variables adopting $\delta
a=-$0.17([Fe/H]+2.1). As for the effects of mass loss, we recall that
the predicted limits become brighter by $\sim$ 0.1 mag with a mass
loss of $\sim$ 0.7\msun.

For all the galaxies, we now find an excellent agreement with the
predicted FOBE and FRE, with very few exceptions: the variables v1 in
AndI, v8 in AndIII and (by a minor extent) J023952.5 in Fornax, which
are fainter than the FRE, and the variables v193 in Carina and v10320
in LMC, wich are slightly brighter than the FOBE. As already shown in
Table 4, in this plane the predicted blue edge for fundamental
pulsators (FBE, dashed line) is almost coincident with the red edge
for first overtone pulsators (FORE, dotted line). Taking into account
the uncertainty of the predicted edges ($\sim$ 0.12 mag) and of the
measured magnitudes ($\sim$ 0.16 mag), we preliminarily assign the
fundamental mode (F) to the variables $\ge$ 0.2 mag fainter than FORE
and the first overtone mode (FO) to those $\ge$ 0.2 mag brighter than
the FBE, leaving the few remaining variables with an uncertain (FFO)
classification. ACs (filled circles) and spCCs (open squares) in LMC
and Phoenix populate a rather common instability strip, with the
latter variables located at brighter magnitudes and longer periods.
Conversely, the P2Cs observed in Fornax have a distinct behavior,
being significantly fainter than the AC fundamental red edge. Such
evidence leads us to suppose that v1 in AndI and v8 in AndIII
might be Population II Cepheids, as already suggested by Pritzl et
al. (2005).

\begin{table}[h]
\begin{center}
\caption{Anomalous Cepheids not fulfilling the predicted edges of the
instability strip into the Color-Magnitude ($CM$), Period-Color ($PC$)
or Period-Magnitude ($PM_V$) diagram. The label ``yes'' means that the
variable agree with the predicted edges in a given diagram, otherwise
we give the reason for the disagreement.} \label{PLCM}
\begin{tabular}{llll}
\hline\hline
gal/var      &  $CM$        &   $PC$   &         $PM_V$      \\
\hline
AndI\\
1&       $B-V<$FOBE         &   yes       &         $M_V>$FRE  \\
AndIII\\
1&   yes                  &       $B-V<$FOBE   &   yes                 \\
8&   yes                  &      yes              &         $M_V>$FRE  \\
Carina\\
193&   yes                  &    yes                &         $M_V<$FOBE \\
Fornax\\
J0+\\
24050.2&          $V-I>$FRE  &       $V-I>$FRE    &   yes               \\
23907.1&          $V-I>$FRE  &       $V-I>$FRE    &    yes              \\
24002.7&          $V-I>$FRE  &       $V-I>$FRE    &   yes               \\
23937.7&          $V-I>$FRE  &       $V-I>$FRE    &   yes               \\
23946.2&          $V-I>$FRE  &       $V-I>$FRE    &   yes               \\
23952.5&          $V-I>$FRE  &       $V-I>$FRE    &         $M_V>$FRE      \\
LMC\\
      10320&      yes               &       $B-V>$FRE    &         $M_V<$FOBE      \\
      9578&          $B-V>$FRE  &       $B-V>$FRE    & yes                 \\
     5952&          $B-V>$FRE  &       $B-V>$FRE    &  yes                \\
Phoenix\\
   103951&          $B-V<$FOBE &       $B-V<$FOBE   &  yes                \\
   6793&          $V-I>$FRE  &       $V-I>$FRE    &    yes              \\
   11219&          $B-V<$FOBE &       $B-V<$FOBE   &    yes               \\
   10800 &                    &                    & yes    \\
Sextans\\
      34&     yes                &       $B-V>$FRE?   &  yes                 \\
      19&     yes                 &       $B-V>$FRE?   &  yes                 \\
\hline
\end{tabular}
\end{center}
\end{table}

\begin{figure}
\includegraphics[width=8cm]{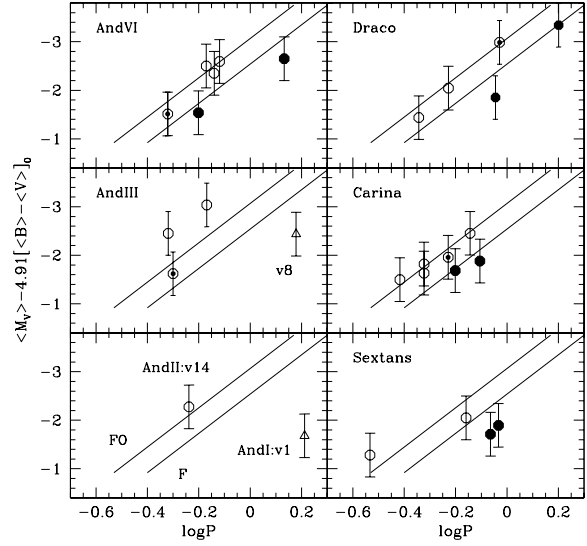}
\caption{\small{Period-Magnitude-Color distribution of observed ACs
compared to the predicted relation for F and FO pulsators, in
the absence of mass loss. Filled and open circles are F and FO
pulsators, respectively, while the circled dots refer to FFO
variables. The open triangles show the suspected P2Cs.}}
\end{figure}

\begin{figure}
\includegraphics[width=8cm]{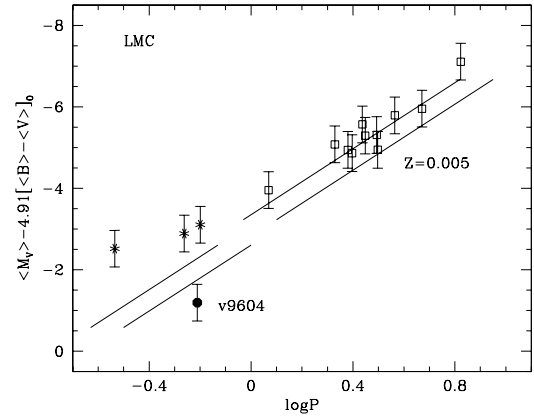}
\caption{\small{As in Fig.  23, but for the variables in LMC. The
asterisks refer to the AC candidates not fullfilling the predicted
instability strip requirements (see Table 10). The open squares show
spCCs.}}
\end{figure}

\begin{figure}
\includegraphics[width=8cm]{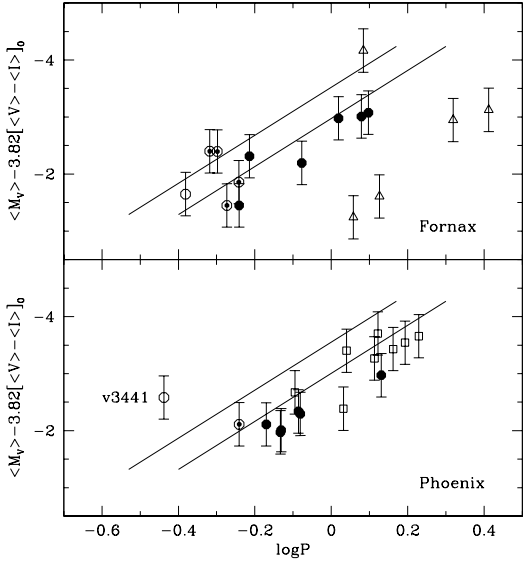}
\caption{\small{As in Fig. 23, but for the observed ACs in Phoenix and
Fornax. We plot also spCCS (open squares) and P2Cs (open triangles).}}
\end{figure}

The results of the comparison between the observed $CM$, $PC$ and
$PM$ distributions and the predicted edges of the whole instability
strip, as summarized in Table 10, show that each of these
bi-dimensional diagrams can yield discordant results. This is not a
surprise since the properties of individual variables are fully
described by a four-dimensional formulation as given by Eq. (1) or,
in the case of variables following the same $ML$ relation, by the
three-dimensional one provided by the $PMC_e$ relation. 

Using only the variables filling the predicted instability strip,
i.e., excluding those listed in Table 10, and adopting the
coefficients given in Table 7 to correct the absolute magnitude for
the color, we show in Figs. 19-21 the comparison with the predicted
$PMC_e$ relations in absence of mass loss. In these figures, the F and
FO pulsators identified in the $PM_V$ diagram are plotted with filled
and open circles, respectively, while the circled dots refer to FFO
pulsators. Moreover, the predicted $PMC_e$ relations, which are given
at $Z$=0.0001 as a function of $P_F$, are corrected for the metal
content adopting $\delta \alpha_0=-$0.17([Fe/H]+2.1) and are also
plotted shifted by $\delta$log$P=-$0.13 to account for FO
pulsators. The errorbars are drawn to account for the photometric
errors [$\epsilon (M_V)=\pm$0.16 mag and
$\epsilon(B-V)=\epsilon(V-I)\pm$0.07 mag] and the intrinsic
uncertainty ($\pm$ 0.08 mag) of the predicted $PMC_e$ relations.

Inspection of the figures yields the following results:

\begin{itemize}
\item {\it AndI}: We can definitely conclude that v1 is a
P2C
\item {\it AndII}: We confirm that v14 is a FO pulsator;
\item {\it AndIII}: The variable v8 is a P2C.
The two FO candidates v6 and v7 are located above
the predicted $PMC_e$ relation, suggesting that the measured $B-V$
color is too red or the adopted distance modulus too long.
In the latter case, a difference $\delta \mu_0\sim-$ 0.3 mag would
yield a statistical agreement, as well as the assignment of the
fundamental mode to the variable v9. Otherwise,
if the measured colors and the adopted distance modulus are correct,
v6 and v7 are not ACs and no safe
pulsation mode can be assigned to v9;
\item {\it AndVI}: All the observed ACs agree with the predicted
relations, but no clear pulsation mode can be assigned to the FFO
variable v93 (classified as F by MFC);
\item {\it Sextans}: There is a statistical agreement with the predicted
relations;
\item {\it Carina}: All the F and FO candidates agree quite well with
the predicted relations. The FFO variable v33 is a FO pulsator, in
agreement with MFC;
\item {\it Draco}: All the F and FO candidates agree with the predicted
relations. The FFO variable v157 is a FO pulsator, as suggested by MFC;
\item {\it LMC}: The three AC candidates not fitting in the predicted
edges of the IS (asterisks) are clearly at odds with the predicted
$PMC_e$ relations, while a marginal agreement is found for the
remaining fundamental candidate v9604. The observed spCCs (open
squares) are fitted once the typical metal content of the
Classical Cepheids in LMC ([Fe/H]=$-$0.4, i.e., $Z$=0.005) is taken
into consideration;
\item {\it Phoenix}: Adopting $V-I$ colors, which give less
noisy results than $B-V$ ones, we derive that all the ACs,
except v3441, would agree with the predicted $PMC_e$ relations if a
slightly larger distance modulus ($\delta \mu_0\sim$ 0.2 mag) {\bf
were} adopted. This is also supported by the observed spCCs (open
squares). In such a case, v12003 is a FO pulsator, whereas the
peculiar location of v3441 above the FO predicted relation would
suggest that the measured color is too red;
\item {\it Fornax}: All the ACs are in reasonable agreement with the
predicted relations, and we can assign the first overtone mode to
J023926.8 and J024058.3, while J023941.5 and J024016.9 are pulsating
in the fundamental mode. The P2Cs (open triangles) are located
at significantly longer periods than the fundamental
$PMC_e$ relation, except J023811.9 whose color [$(V-I)_0$=1.06 mag) is
more than 0.4 mag redder than the other P2Cs.
\end{itemize}

\begin{figure}
\includegraphics[width=8cm]{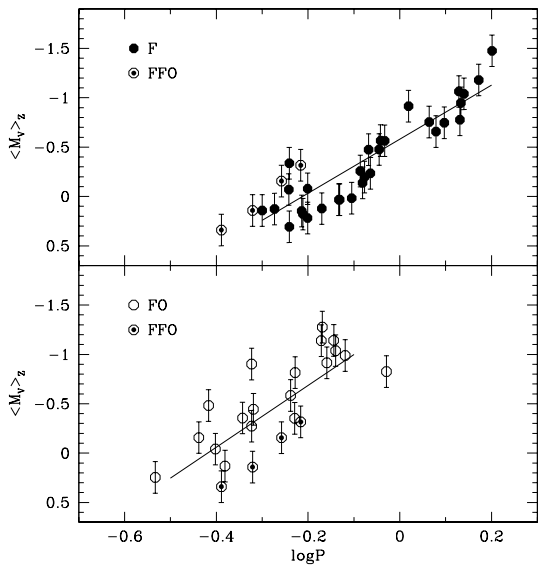}
\caption{\small{Safe AC candidates in the Period-Magnitude diagram, with
the visual magnitude corrected for the metal content. The solid lines are the best
fit to the data.}}
\end{figure}

Similar results are obtained in the case of mass loss, i.e., by using
the predicted $PMC_e$ relations given in Table 8. The AC candidates
which satisfy the predicted edges of the IS in the $CM$, $PC$ {\it
and} $PM_V$ diagram can be used to obtain empirical $PM_V$ relations.
This is shown in Fig.22, where $\langle M_V\rangle_Z$=$\langle
M_V\rangle$+0.17([Fe/H]+2.1) and the solid lines are the linear
regression through the points
$$M_{V,Z}^F=-0.58-2.74\log P_F$$

$$M_{V,Z}^{FO}=-1.31-3.13\log P_{FO}$$

\noindent as derived giving to the FFO pulsators a half weight with
respect to {\bf safe} F and FO variables. These relations are in good agreement
with the ones presented in MFC for a smaller sample of pulsators and
neglecting the metallicity effect.

\section{Conclusions}

We have presented an homogeneous and updated theoretical scenario
for the study of AC properties based on a new set of
evolutionary tracks with $Z$=0.0001, progenitor mass in the range
1.3-2 \msun and  several assumptions on the efficiency of mass loss.
This evolutionary framework is related to the pulsation model
predictions by MFC. We find that predicted pulsators with mass
1.3-2.0 $M_{\odot}$ follow a ML relation as given by
$\langle$log$L\rangle \sim$ 1.77($\pm$0.05)+2.07log$M$, in absence
of mass loss. Moreover, the evolutionary and pulsational models have
been used to build synthetic populations of AC pulsators, each
characterized by a progenitor mass, a mean mass and a percentage of
central He-burning stars populating the IS. We find
that  pulsators with a given mass originating in the presence of mass
loss are systematically fainter than the ones in the absence of mass
loss, with the decrease in luminosity following the variation in the
He-core mass. We have also investigated the effect of mass loss on
the predicted IS boundaries in the $CM$, $PC$ and $PM$ planes and we
find that the only significant dependence occurs  in the $PM$ plane
where the synthetic distribution in the presence of mass loss is, on
average, brighter by about 0.1 mag than the one in the absence of
mass loss.

Concerning the existence of a $PM$ relation for these stars, we
confirm that in the case of optical magnitudes it depends on the
pulsator distribution within the IS, whereas tight near-infrared
$PM_K$ relations can be derived for both fundamental and first
overtone pulsators, providing a reliable tool for distance evaluations
with an intrinsic uncertainty of about 0.15 mag. If the mass term is
taken into account a tighter (mass-dependent) PM$_K$ relation
(r.m.s=0.04 mag.) is obtained. Conversely, the predicted $PC$
relations dealing with $B-V$ and $V-I$ colors appear useful for
reddening determinations with an intrinsic uncertainty of less
than 0.1 mag.  Mass-dependent $PMC$ relations have been derived by
using all the models, as well as evolutionary $PMC_e$ relations where
the mass-term is removed. The latter relations allow distance
determinations with a formal uncertainty of the order of 0.1 mag,
provided that the intrinsic colors are well known.

To take into account the metallicity effect, we have used other
evolutionary tracks from the literature with $Z$ in the range
0.0001-0.001. In particular, we find that an increase of the metal
content from 0.0001 to 0.0004 yields that, at fixed mass, the
average luminosity of predicted AC pulsators decreases by $\sim$ 0.1
dex, producing an effect on the IS boundaries similar to the one due
to mass loss, whereas the zero-points of the predicted $PMC_e$
relations becomes brighter  by $\sim$ 0.1 mag.

By comparing the predicted edges of the IS with observed pulsators in
the $CM$, $PC$ and $PM$ planes, we find discordant results in the
properties of individual stars, thus confirming that a sure
identification of actual ACs requires simultaneous information on
period, magnitude and color through the application of the predicted
$PMC_e$ relations. In this case, we are also able to provide
constraints on the pulsation mode.  We show that AC candidates that
satisfy the predicted edges of the IS in the $CM$, $PC$ and $PM_V$
diagrams can be used to obtain empirical $PM_V$ relations for F and FO
pulsators.

\begin{acknowledgements}
Financial support for this study was provided by MIUR, under the
scientific project ``Continuity and Discontinuity in the Milky Way
Formation'' (PI: Raffaele Gratton).
\end{acknowledgements}



\end{document}